\documentstyle[aps,twocolumn,epsf]{revtex}

\begin{document}

\title{A two atomic species superfluid}

\author{G. Modugno, M. Modugno, F. Riboli, G. Roati\thanks{also at Dipartimento di Fisica, Universit\`a di
    Trento, 38050 Povo, Italy}, and M. Inguscio}

\address{European Laboratory for Nonlinear Spectroscopy and Dipartimento di Fisica, Universit\`a di Firenze, and INFM,
 Via Nello Carrara 1, I-50019 Sesto Fiorentino, Italy }

\date{\today} 

\maketitle

\begin{abstract}

We produce a quantum degenerate mixture composed by two Bose-Einstein condensates of different atomic species, $^{41}$K and $^{87}$Rb. We study the dynamics of the superfluid system in an elongated magnetic trap, where off-axis collisions between the two interacting condensates induce scissors-like oscillation. 
\end{abstract}

\pacs{03.75.Fi, 05.30.Jp, 67.90.+z}

The long-standing interest in mixtures of superfluids, originally focused on 
helium systems \cite{helium}, has recently been renewed by the achievement of Bose Einstein 
condensation (BEC) in dilute atomic gases \cite{ketterle}. Already using a single atomic species, multiple condensates were realized by exploiting the magnetic 
structure of the ground electronic state of alkali atoms. Mixtures of two 
hyperfine spin states of $^{87}$Rb in magnetic traps allowed to study the effect of 
the mutual interaction in the dynamic of miscible BECs \cite{rbbec}. Superposition 
of spinor condensates of $^{23}$Na in an optical trap led to a first observation 
of both weakly miscible and immiscible superfluids \cite{ketterle1} and of the occurrence 
of metastable states \cite{ketterle2}. These experimental achievements stimulated an extensive theoretical research on the properties of a mixture of two BECs, and the role of the interparticle interaction in determining its static and dynamical properties has been recognized \cite{topology,stability,phase,dynamics,riboli}.\par
As early suggested \cite{topology}, an even wider scenario for the study of superfluid systems would be opened by BEC in mixtures of different atoms. Considering the species condensed so far, the use of different isotopes of the same species would be restricted to the case of rubidium \cite{isotopes}, while a wider choice would be offered the use of different atomic species. Recently, two-species mixtures were successful for the realization of Fermi-Bose degenerate gases \cite{fb}.\par
In this Letter, we report the realization of a mixture of Bose-Einstein condensates of different atomic species, using potassium and rubidium. Simultaneous condensation of $^{41}$K and $^{87}$Rb is achieved by means of two-species sympathetic cooling \cite{science} in a magnetic trap. The stability against collapse of the degenerate mixture, already forecast from the repulsive character of the strong interspecies interaction \cite{coll}, makes the system suitable for a large variety of investigations. In particular, we explore in this work the dynamics of the two interacting condensates in the magnetic trap, and we observe scissors-like oscillations induced by off-axis collisions. \par
The production of the binary BEC (or TBEC) is based on the experimental apparatus described previously \cite{science}. In brief, $^{41}$K atoms at 300~$\mu$K and $^{87}$Rb  at 
100~$\mu$K are loaded in a QUIC magnetostatic trap \cite{esslinger} using a double magneto-optical trap apparatus. 
Both species are prepared in their $|$F=2, M$_F$=2$\rangle$ state, and they experience the same trapping potential with cylindrical symmetry. The axial and radial harmonic frequencies for Rb are $\omega_{a}$=2$\pi\times$ 16.3~Hz and $\omega_{r}$=2$\pi\times$190~Hz respectively, while those for K are a factor
$\sqrt{M_{Rb}/M_K}$=1.45 larger. Evaporative cooling is performed selectively 
on the Rb sample using a microwave knife tuned to the hyperfine transition at 
6.8~GHz, while the K sample is sympathetically cooled through elastic K-Rb 
collisions. We have slightly but significantly changed the evaporation strategy with respect to that reported in Ref.~\cite{science} by adding a second microwave knife to remove Rb atoms from $|$F=2,M$_F$=1$\rangle$. Indeed, we found that even a small, residual fraction of atoms in such state after the optical pumping phase can cause relevant losses on K, due to the relatively large collisional rate \cite{coll}. Using this strategy we are now able to cool down to 
condensation about 10$^{4}$ atoms of each species in 50~s, starting from 
10$^{5}$ potassium and 5$\times$10$^{8}$ rubidium atoms.\par

\begin{figure}[htb]
\begin{center}
\leavevmode 
\epsfxsize=7cm \epsfbox{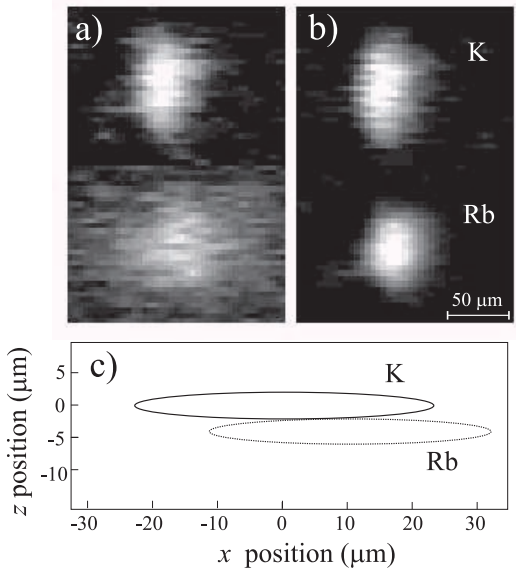} 
\end{center}
\caption{Absorption imaging of a binary $^{41}$K-$^{87}$Rb Bose-Einstein condensate after 13~ms of ballistic expansion: a) at $T\simeq$100~nK the Rb sample is still thermal; b) at $T\leq$80~nK both species are condensed. c) Enlarged view of the density profile of the binary BEC in the magnetic trap.} 
\label{fig1} 
\end{figure}

Since at this stage the number $N$ of K and Rb atoms is comparable, their critical temperatures for BEC scale only with the atomic mass $M$ as $T_C$=$\hbar/k_B (\omega_{r}^2\omega_{a} 
N/1.2)^{1/3}$ $\propto$ $M^{-1/2}$. Therefore, as shown Fig.~\ref{fig1}, condensation is reached first for K at a temperature of about 120~nK (Fig.~\ref{fig1}a), and then for Rb at 80~nK (Fig.~\ref{fig1}b). The two BECs are 
probed simultaneously at the end of each experimental run by absorption 
imaging, after about 13~ms of ballistic expansion.  The two condensate appear to be separated 
along the vertical direction because of the imaging procedure. Indeed, K and 
Rb atoms are imaged on two different regions of a charge-coupled-device (CCD) 
camera, using two 30-$\mu$s light pulses delayed by 700~$\mu$s, at a 
wavelength of 767.9~nm and 780~nm respectively. Because of the different trap frequencies, the aspect ratios for the K and Rb BECs are different at the same expansion time.\par
 By taking various images of the two condensates at different expansion times, we could reconstruct their size and relative position in the magnetic trap. The experimental observation is in good agreement with the result of a general analytical model for the ground state of a TBEC \cite{riboli}, for our parameters; the simulated profile is shown in Fig.~\ref{fig1}c. The radii of the K (Rb) BEC are R$_x$=23~$\mu$m (22~$\mu$m) and R$_z$=2~$\mu$m (1.9~$\mu$m). The two centers are separated, due to the gravity, in both in the radial and axial directions ($\delta z$=3.6~$\mu$m, $\delta x$=10~$\mu$m); the sag along the weak 
$x$-axis is caused by a small tilt of the trap axes with respect to 
gravity, by an angle $\alpha\simeq
$20~mRad.\par

We measure temperatures from the thermal gaussian backgrounds, for which we have a detection limit of 30\%. This implies that for each condensate we can directly measure the temperature down to $T$=0.65$T_C$. As a matter of fact, the critical temperature for BEC of Rb is also the lowest temperature measured for K (0.65$\times$120~nK $\simeq$ 80~nK). This is an evidence of the thermalization between the two species at the onset of quantum degeneracy. Although the region of overlap of the two BECs in our magnetic trap is very small, in this regime the interspecies thermalization can be mediated by the thermal clouds, which are instead partially overlapping even at the lowest temperatures. Our observations are suggestive of an efficient sympathetic cooling  mechanism, also considering that in the temperature interval 80$\div$120~nK the heat capacity of the K sample is increased with respect to its classical value \cite{dalfovo}.\par
The lifetime of the TBEC is the same as for a single Rb BEC in our 
system, and it is limited to about 500~ms by the background heating of the magnetic trap. 
The stability of the mixture confirms the repulsive character of the $^{41}$K-
$^{87}$Rb interaction, which we previously determined by collisional measurements on 
thermal samples to be $a$=163$^{+60}_{-15}~a_0$ \cite{coll}. Within the mean field approach the atom-atom interaction strengths which characterize the properties of a TBEC are
$g_{ij}$=$2\pi\hbar^2 a_{ij}/\mu_{ij}$ where the suffixes $i$ and $j$ enumerate the components, a$_{ij}$ are the relevant $s$-wave scattering lengths, and $\mu_{ij}$ is the reduced mass for two atoms of species $i$ and $j$. It has been shown \cite{stability,riboli} that the stability of a mixture 
of two BECs which are individually stable ($g_{11}>$0, $g_{22}>$0) depends on the 
value of the quantity $\Delta$=$g_{12}/\sqrt{g_{11}g_{22}}$. In particular, if 
$\Delta<$-1 a binary BEC would not exist, because the mean-field attraction  
between atoms of the two distinct condensate would overwhelm the 
corresponding repulsion between atoms of the same specie, leading to a collapse. In our case, the intraspecies triplet scattering lengths are $a_{11}$=60~$a_0$ \cite{kscatl} and $a_{22}=99~a_0$ \cite{rbscatl} for $^{41}$K and $^{87}$Rb, respectively. A negative $a_{12}$ which would lead to $\Delta$=-3 and would imply the collapse of the TBEC, is therefore ruled out by our observation.\par
\vspace{0.5cm}
\begin{figure}[htb]
\begin{center}
\leavevmode 
\epsfxsize=7.5cm \epsfbox{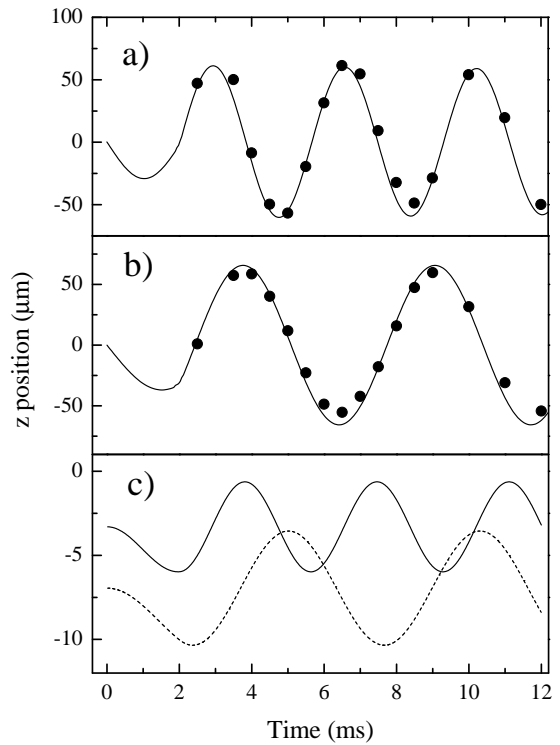} 
\end{center}
\caption{Dipolar oscillations of the K and Rb condensates along the $z$ axis, induced by a modification of the radial trap confinement. Center-of mass position of a) K and b) Rb after the ballistic expansion; the dots are the experimental data, while the continuos lines are a simulation using the GPE. c) Calculated evolution of the center-of-mass position for K (continuous line) and Rb (dotted line) in the trap.} 
\label{fig2} 
\end{figure}

Because of gravity, the two BECs are almost completely separated in the magnetic trap. We therefore forced the two condensates to overlap and to interact by inducing dipolar oscillations. A good degree of overlap is obtained with small amplitude oscillations along the vertical trap axis, forced by a reduction of the radial confinement by about 25\% for 2~ms. In Fig.~\ref{fig2}a-b we report the evolution of the center-of-mass positions of the two BECs after the ballistic expansion, during the first few oscillations. This measurement has been performed with a TBEC composed by typically 6$\times 10^{3}$ K atoms and $10^{4}$Rb atoms, with no detectable thermal fractions. The experimental data are well reproduced by the numerical solution of the time-dependent Gross-Pitaevskii equations (GPE) for our TBEC (continuos lines in Fig.~\ref{fig2}a-b). The simulation also allows to reconstruct the corresponding motion in the magnetic trap, which is shown in Fig.~\ref{fig2}c. Due to the different trap frequencies for K and Rb, the oscillations of the two BECs get rapidly out of phase after the reduction of vertical confinement, and the two BECs can repeatedly collide; the amplitude of the oscillations is such to produce situations of complete geometrical overlap of the two samples along the vertical direction.\par
\begin{figure} 
\begin{center}
\leavevmode 
\epsfxsize=9cm \epsfbox{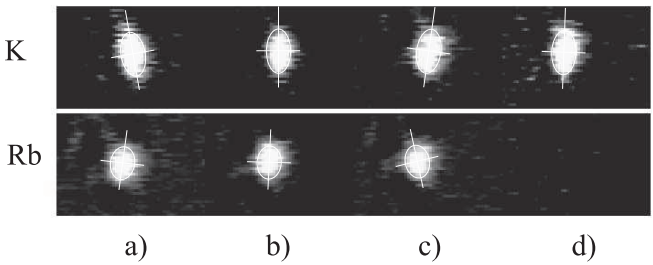} 
\end{center}
\caption{Collision-induced rotation of the binary BEC during the dipolar oscillations. The first three images correspond to an evolution time in the trap: a) 8~ms; b) 8.5~ms; c) 9~ms. In d) the evolution time is 8~ms like in a), but the Rb BEC has been removed before the excitation of the dipolar oscillations; no rotation is observed in this case. } 
\label{fig3} 
\end{figure}

The experimental observation indicates that the mutual repulsion of the two condensates does not strongly affect their center-of-mass motion on the timescale of a few oscillations \cite{damping}. However, a clear effect of the interaction appears in a study of the shape of the TBEC. Taking advantage of the fact that we can selectively remove either one of the two species from the trap by means of a proper light pulse, we can study the oscillations when only one or both condensates are present. In the latter case, we observe a time-dependent rotation of the two condensates, as shown in Fig.~\ref{fig3}a-c. The time-dependent rotation is more evident for K, due to the larger aspect ratio. When instead Rb is expelled from the trap just after the formation of the TBEC, so that the dipolar dynamics involves only the K BEC without collisions with Rb, no rotation of the symmetry axis is observed, as shown in Fig.~\ref{fig3}d. The rotation is caused by an exchange of angular momentum between the two BECs during the collisions, as a consequence of the displacement $\delta x$ of the two centers of mass along the weak trap axis. Indeed, in the mean-field approach each BEC feels the presence of the other one as a time-dependent modification of its own trapping potential. Since the axial symmetry of the effective potential is broken, the BEC acquires a macroscopic angular momentum. \par
In contrast, two classical gases would behave in a qualitatively different way during the same dipolar oscillations, since their dynamics would be determined by uncorrelated two-body collisions. In particular, a rotation like a rigid body would take place for classical gases only in the {\it collisional} hydrodynamic regime, where the collisional rate is larger than the mean trap frequency. However, in this regime the center-of-mass motion would be damped on the timescale of the trap period, differently from what we observe for the two condensates, whose motion is governed by the hydrodynamic equations for superfluids in the {\it collisionless} regime.\par
\vspace{0.5cm} 
\begin{figure} 
\begin{center}
\leavevmode 
\epsfxsize=8.5cm \epsfbox{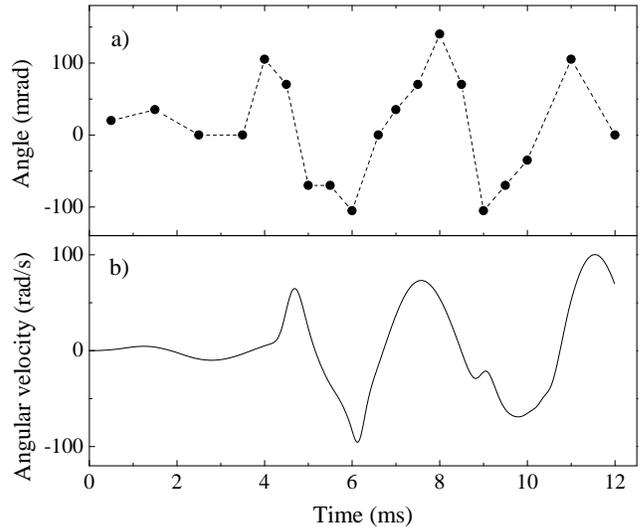} 
\end{center}
\caption{a) Evolution of the angle of rotation of the K BEC from the vertical direction, after a 13~ms ballistic expansion. The dotted line is a guide to the eye. b) Numerical simulation using the GPE of the angular velocity of the condensate at release from the trap.} 
\label{fig4} 
\end{figure}

The superfluid nature of the system is evidenced also by the peculiar behavior of the rotating BECs during the ballistic expansion. Indeed, the exchange of angular momentum between the two BECs results in small rotation angles between the long axis of each BEC and the $x$ axis of the magnetic trap, while they are trapped. Once the elongated condensates are released and stop interacting between themselves, their rotation angle evolves in a non-classical way, as discussed in Ref.~\cite{stringari}. Following that analysis, as soon as the aspect ratio gets close to unity the irrotationality of the velocity field within the condensates forces their initial horizontal axis to rotate much faster and to approach the vertical direction. The total angle described by the condensates during the expansion depends on the initial angular velocity, and is smaller than $\pi$/2. For a non-rotating BEC the unity aspect ratio is reached after an expansion time $\tau\simeq1/\omega_a$. In our experiment, $\tau$ is approximately 7~ms and 10~ms for K and Rb, respectively, in accordance with the observation of small angles from the $z$ direction for K and larger angles for Rb. We have actually verified by solving the Thomas-Fermi hydrodynamic equations \cite{stringari} that for our expansion time of 13-ms the K BEC is already close to its asymptotic angle, while the Rb BEC is still in the region of large angular velocity.\par
We have studied the oscillations of the angle of the long axis of the K BEC from vertical, as a function of the dwelling time in the trap, at a fixed expansion time of 13~ms. The results are reported in Fig.~\ref{fig4}a. The condensate axis starts to oscillate after about 4~ms, when the first collision occurs, and the frequency of the induced oscillation is close to $\omega_r$. However, a pure sinusoidal scissors mode \cite{scissor} cannot occur, since the two BECs collide periodically. A simulation using the time-dependent GPE for the evolution of K-Rb TBEC in the trap confirms the issue of a collisional-induced rotation for our experimental parameters. From the simulation we can deduce the angle and the angular velocity of the K BEC at the release from the trap. We discover that the measured angle after the expansion is proportional to the angular velocity at release, as reported in Fig.~\ref{fig4}b. This is a remarkable behavior, which is suggestive of important implications also for a pure scissor oscillation of an expanding BEC.\par

In conclusion, we have produced a binary BEC composed by two different atomic species, and we have induced scissors-like oscillations by means of interspecies collisions.  The observed phenomenology evidences the superfluid nature of the system, and opens new directions for the study of scissors oscillations. The novel system is also likely to open new possibilities for the investigation of phase separation between two degenerate gases, because of the large repulsive interaction. Symmetry-breaking phenomena and metastability effects \cite{topology} could be investigated, for instance in combination with optical trapping. Regarding to the process of sympathetic cooling, a new system is now available for the study of the subtle issue of thermalization at the onset of quantum degeneracy.\par

Finally, the K-Rb TBEC should play also an important role for the formation of ultracold 
heteronuclear molecules, also in combination with photoassociative schemes
\cite{photoass}, with magnetically-tunable Feshbach resonances \cite{fesh}, or a combination of the two \cite{mackie}. Ultracold dipolar bosons would allow to study trapped degenerate gases with long-range interactions \cite{dipolar}, or even to implement quantum computing schemes \cite{demille}.\par

We acknowledge useful discussions with A. Simoni and S. Stringari. This work was supported by MIUR, by ECC under the Contract HPRICT1999-00111, and by INFM, Progetto di Ricerca Avanzata 
``Photonmatter".


\begin{thebibliography}{99}
\bibitem{helium}L. Guttman and J. R. Arnold, Phys. Rev. {\bf 92}, 547 (1953); Khalatnikov, Sov. Phys. JETP {\bf 5}, 542 (1957); W. B. Colson and A. L. Fetter, J. Low. Temp. Phys. {\bf 33}, 231 (1978).
\bibitem{ketterle} For a review, see J. R. Anglin and W. Ketterle, Nature {\bf 416}, 
212 (2002).
\bibitem{rbbec} C. J. Myatt, {\it et al.}, Phys. Rev. Lett. {\bf 78}, 586 (1997); D. S. Hall, {\it et al.}, Phys. Rev. Lett. {\bf 81}, 1539 (1998); 
P. Maddaloni, {\it et al.}, Phys. Rev. Lett. {\bf 85}, 2413 (2000).
\bibitem{ketterle1} J. Stenger, {\it et al.}, Nature {\bf 396}, 345 
(1998). 
\bibitem{ketterle2} H.-J. Miesner, {\it et al.},  Phys. Rev. Lett. {\bf 82},
2228 (1999).
\bibitem{topology}T. L. Ho and V. B. Shenoy, Phys. Rev. Lett. {\bf 77}, 3276
(1996); H. Pu and N. P. Bigelow, Phys. Rev. Lett. {\bf 80}, 1130
(1998).
\bibitem{stability} B. D. Esry, {\it et al.}, Phys. Rev. Lett. {\bf 78}, 3594
(1997); C. K. Law, {\it et al.}, Phys. Rev. Lett. {\bf 79}, 3105 (1997).
\bibitem{phase} E. Timmermans, Phys. Rev. Lett. {\bf 81}, 5718 (1998); P. Ao and S. T. Chui, Phys. Rev. A {\bf 58}, 4836 (1998). R. Barankov, cond-mat/0112326.
\bibitem{dynamics} Th. Bush, {\it et al.}, Phys. Rev. A {\bf 56}, 2978 (1997);
R. Graham and D. Walls, Phys. Rev. A {\bf 57}, 484 (1998); H. Pu and N. P. Bigelow, Phys. Rev. Lett. {\bf 80}, 1134 (1998).
\bibitem{riboli} F. Riboli and M. Modugno, cond-mat/0202344.
\bibitem{isotopes} J. P. Burke, {\it et al.}, Phys. Rev. Lett. {\bf 80}, 2097 (1998).
\bibitem{fb} Z. Hadzibabic, {\it et al.}, Phys. Rev. Lett. {\bf 88}, 160401
(2002); G. Roati, {\it et al.}, cond-mat/0205015.
\bibitem{science} G. Modugno, {\it et al.}, Science {\bf 294}, 1320 
(2001).  
\bibitem{esslinger} T.  Esslinger, I. Bloch, T.W. H\"ansch, Phys. Rev. A
{\bf 58}, R2664 (1998).
\bibitem{coll} G. Ferrari, {\it et al.}, cond-mat/0202290.
\bibitem{dalfovo}F. Dalfovo, {\it et al.},  Rev. Mod. Phys {\bf 71},
463 (1999).
\bibitem{kscatl} H. Wang, {\it et al.}, Phys. Rev. A {\bf 62}, 052704 (2000).
\bibitem{rbscatl} E.G.M. van Kempen, {\it et al.}, Phys. Rev. Lett. {\bf 88},
093201 (2002).
\bibitem{damping} Damping and coupling of the two motions are observed on a timescale longer than 100~ms, in good agreement with the solutions of the GPE. 
\bibitem{stringari} M. Edwards, {\it et al.},  Phys. Rev. Lett. {\bf 88},
070405 (2002).
\bibitem{scissor} D. Guery-Odelin and S. Stringari, Phys. Rev. Lett. {\bf 83},
4452 (1999); O. M. Marag\`o, {\it et al.},  Phys. Rev. Lett. {\bf 84},
2056 (2000).
\bibitem{photoass} P. D. Drummond, K. V. Kheruntsyan, and H. He, Phys. Rev. Lett. {\bf 81}, 3055 (1998).
\bibitem{fesh} E. Timmermans, {\it et al.},  Phys. Rep. {\bf 315},
199 (1999).
\bibitem{mackie} M. Mackie, cond-mat/0202041.
\bibitem{dipolar} K. G\'oral, L. Santos, M. Lewenstein, Phys. Rev. Lett. {\bf 88}, 170406 (2002).
\bibitem{demille} D. De Mille, Phys. Rev. Lett. {\bf 88}, 067901 (2002).


\end{thebibliography}
\end{document}